\documentclass[%
 reprint,prl,
superscriptaddress,
showpacs,
 amsmath,amssymb,
 aps,
]{revtex4-1}

\usepackage{graphicx}% Include figure files
\usepackage{dcolumn}% Align table columns on decimal point
\usepackage{bm}% bold math
\usepackage{hyperref}% add hypertext capabilities
\usepackage{verbatim} 
\usepackage{color}
\usepackage{pifont}
\usepackage{bbold}
\usepackage[normalem]{ulem}

\usepackage{pdfpages}
\usepackage{lipsum}

\newcommand{\tint}{\tau_{\mathrm{int}}}

\newcommand{\Dthresh}{D_\text{crit}}
\newcommand{\Dcl}{D_\text{cl}}
\newcommand{\tcollision}{\tau_{\mathrm{c}}}
\newcommand{\tgrowth}{\tau_{\mathrm{g}}}
\newcommand{\tlag}{\tau_{\mathrm{\ell}}}

\newcommand{\average}[1]{\langle #1 \rangle}
\newcommand{\hot}{\textnormal{hot}}
\newcommand{\cold}{\textnormal{cold}}

\newcommand{\Nsurface}{N_\sigma}

\newcommand{\Minfty}{M_{\infty}}

\newcommand{\vek}[1]{{\bf #1}}

\newcommand{\asqDhot}{a^2/\Dhot}

\newcommand{\alphap}{\alpha_{\textnormal{att}}}
\newcommand{\alpham}{\alpha_{\textnormal{coll}}}
\newcommand{\alphac}{\alpha_{\textnormal{int}}}

\newcommand{\rijhat}{\hat{\vek{r}}_{ij}}
\newcommand{\Fij}{\vek F_{ij}}

\newcommand{\Deff}{D_{\textnormal{eff}}}

\newcommand{\PackingFractionHot}{\phi_{\hot}}

\newcommand{\Dhot}{D_{\hot}}
\newcommand{\Dcold}{D_{\cold}}
\newcommand{\Nhot}{N_{\hot}}
\newcommand{\Ncold}{N_{\cold}}

\newcommand{\ch}[1]{\textcolor{black}{#1}}

\usepackage{xr}
\begin{document}

\title{Binary Mixtures of Particles with Different Diffusivities Demix}

\author{Simon N.\ Weber}
\altaffiliation{Present address: Technische Universit\"at Berlin, Marchstr.\ 23, 10587 Berlin, weber@tu-berlin.de}
\affiliation{%
Arnold Sommerfeld Center for Theoretical Physics and Center for NanoScience, \\
Department of Physics, Ludwig-Maximilians-Universit\"at M\"unchen, 80333 Munich, Germany
}

\author{Christoph A.\ Weber}
\affiliation{Max Planck Institute for the Physics of Complex Systems, N\"othnitzer Str.\ 38, 01187 Dresden, Germany}
\author{Erwin Frey}%
\altaffiliation{Correspondence please to frey@lmu.de.}
\affiliation{%
Arnold Sommerfeld Center for Theoretical Physics and Center for NanoScience, \\
Department of Physics, Ludwig-Maximilians-Universit\"at M\"unchen, 80333 Munich, Germany
}%%
             
\begin{abstract}
The influence of size differences, shape, mass and persistent motion on phase separation in binary mixtures has been intensively studied. Here we focus on the exclusive role of diffusivity differences in binary mixtures of equal-sized particles. We find an effective attraction between the less diffusive particles, which are essentially caged in the surrounding species with the higher diffusion constant. This effect leads to phase separation for systems above a critical size: A single close-packed cluster made up of the less diffusive species emerges. Experiments for testing of our predictions are outlined.
\end{abstract}

\maketitle

Binary mixtures consisting of non-adhesive particles in thermal equilibrium can demix if their constituents differ in size~\cite{Asakura:1958tn, Biben:1999ka,Gotzelmann:1998un, Dijkstra:1994bh, Mao_cates_199510, PhysRevE.61.4095, loewen_depletion, Marenduzzo:2006ch}. This demixing is explained in terms of an increase in the entropy of the system. 
However, non-adhesive binary mixtures that are far from thermal equilibrium may demix even when the particles are of equal size.
Examples include mixtures of active and Brownian particles~\cite{McCandlish:2011ix, stenhammar2015activity}, and shaken granular media consisting of particles of different masses~\cite{Rivas:2011ht,Rivas:2011}. 
While the impact of  activity on the phase separation dynamics
has been frequently addressed~\cite{McCandlish:2011ix, stenhammar2015activity}, 
 the specific role of differences in the diffusion constants has not yet been investigated.

\begin{figure}[htb]
\centering
\includegraphics[width=0.9\linewidth]{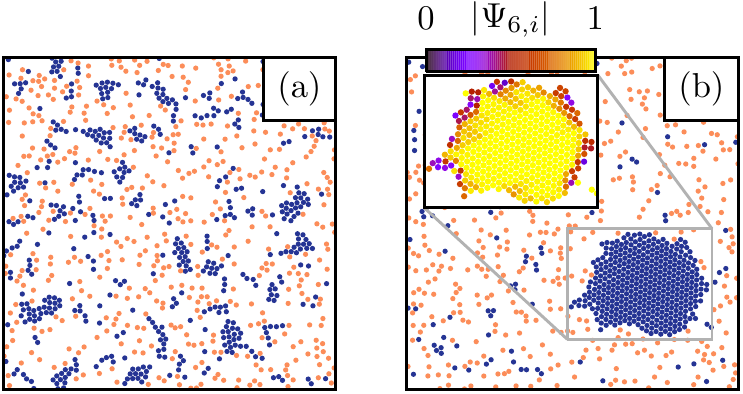}
\caption{%
Snapshots of particle configurations in a binary mixture of 500 hot particles (orange, light gray) and 500 cold particles (blue, dark gray) with diffusion constants that differ by a factor $D=\Dcold/\Dhot=10^{-3}$ for a packing fraction $\phi=0.2$ at \text{(a)} early times $t=6 \cdot 10^3\, \asqDhot$, and \text{(b)} late times $t=6.4\cdot10^4$ $\asqDhot$ of the coarsening dynamics. Initially, small shape-less clusters of cold particles form, which later coalesce and merge into a single cluster.
Inset: Close-up of cluster with color coded local hexatic order parameter $\Psi_{6,i}=|\mathcal{N}_i|^{-1} \sum_{j\in{}\mathcal{N}_i}e^{ \imath 6\theta_{ij}}$ with $\mathcal{N}_i$ denoting the Voronoi neighbors of particle $i$ and $\theta_{ij}$ is the `bond'-angle between particles $i$ and $j$.
}
\label{fig:eye_catcher}
\end{figure}

Here, we study conditions under which diffusivity differences among non-adhesive spherical particles of equal size and shape can lead to demixing.
 We investigate binary mixtures of two particle species which differ only with respect to their diffusion constants, and interact by a short-ranged repulsive force. We find that demixing is promoted for large relative differences in the diffusion constants and high overall packing fractions.
The binary mixture exhibits phase separation into a solid-like cluster of the species with the lower diffusion constant (`cold' particles) and a gaseous phase of the `hot' particles with the higher diffusivity; see Fig.~\ref{fig:eye_catcher}. 
As discussed below, these predictions may be tested using various experimental systems including mixtures of granular disks~\cite{Melby:2005gh, Rivas:2011, Rivas:2011ht, Deseigne:2011thesis} or mixtures of photo-activated colloids~\cite{palacci2013living,Buttinoni:2013_sp}.

The Brownian dynamics of the binary system of $\Nhot$ hot and $\Ncold$ cold particles is described by a set of coupled Langevin equations:
$\partial_t \vek r_i 
= 	\mu \sum_{j=1}^N \Fij(t) + \boldsymbol{\eta}_i(t)$, where the sum runs over all particles $i \in \{1, \ldots, N \}$ with $N= \Nhot+\Ncold$. All particles have the same mobility $\mu$ but differ in the value of their diffusion constants, $D_i \in \{\Dhot, \Dcold \}$, which determine the respective magnitude of the spatially isotropic and Gaussian white noise $\average{\eta_{i \alpha}(t) \eta_{j \beta} (t')} = 2 D_i \delta_{ij} \delta_{\alpha \beta} \delta(t-t')$. 
The interactions between the spherical particles of radius $a$ are taken as short-ranged harmonic repulsive forces, $\Fij = k(2a- r_{ij}) \rijhat$ if particles overlap ($r_{ij} < 2 a$), and $\Fij = 0$ otherwise.  Here, $k$ denotes the spring constant and $r_{ij}=| \vek r_i - \vek r_j |$ the inter-particle distance ($\rijhat = (\vek r_i - \vek r_j)/ r_{ij}$). In order to mimic hard particles, we use large values of $k \cdot \mu $, thus ensuring that particle overlaps decay quickly \ch{(see Supplemental Material~\cite{SM})}. 
Recently, a similar model has been used to study clustering near a hard wall~\cite{awazu2014segregation}.
For specificity, we consider $\Nhot = \Ncold$; we have checked that changing the ratio of hot to cold particles by a factor of two in either direction does not lead to qualitatively different results. As control parameters we consider the ratio of diffusion constants $D := \Dcold/\Dhot$ and the packing fraction $\phi := N \pi a^2/L^2$, where $L$ is the system size of the simulation box with periodic boundary conditions.

Our Brownian dynamics simulations show that for sufficiently large relative differences in the diffusion constants an initially random distribution of hot and cold particles spontaneously segregates into a solid-like phase of cold particles surrounded by a gas-like phase of mainly hot particles; see Fig.~\ref{fig:eye_catcher}(a,b) and Supplemental Material for a video~\cite{SM}.
This clustering instability is consistent with analytic predictions based on a low density expansion
for the same system~\cite{joanny_low_density}.
In our simulations the phase separation process begins with the formation of small, shapeless clusters of cold particles of varying size.
We observe coarsening dynamics as these clusters both coalesce and continue to incorporate individual particles, finally leading to a single, spherical cluster at large time scales. The bulk of this cluster exhibits hexagonal order [inset of Fig.~\ref{fig:eye_catcher}(b) and Supplemental Material \cite{SM} Fig.~S8].
\begin{figure}[b]
\centering
\includegraphics[width=0.8\columnwidth]{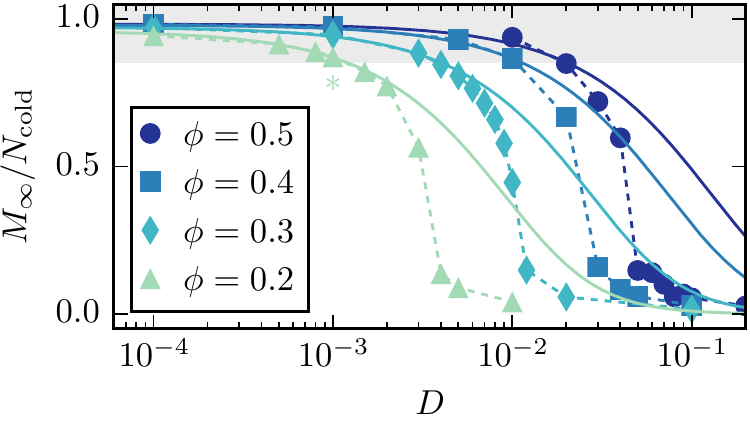}
\caption{%
Fraction of cold particles in the largest cluster of the system, averaged after saturation $\Minfty / \Ncold$ (symbols)  as a function of the diffusion constant ratio $D = \Dcold/\Dhot$ of cold and hot particles for systems with $\Ncold=\Nhot=300$ particles.
Different colors correspond to different packing fractions $\phi$.
Clustering is also observed for packing fractions higher as depicted here (see Supplemental Material~\cite{SM} Fig.~S6).
Values $\Minfty / \Ncold \gtrsim 0.8$ (shaded region) indicate that a single large and stable cluster of cold particles has formed.
The saturation value $\Minfty$ was obtained from 10 realizations.
Solid lines show solutions of Eq.~\eqref{eq:rate_equation} for $t \to \infty$, where the parameter $\alphac$ was fitted to a single set of $D$ and $\phi$ (indicated by $*$); see main text and Supplemental Material~\cite{SM}.}
\label{fig:D_tilde_triggered}
\end{figure} 
Moreover, we have measured the number of cold particles in the largest cluster, $\Minfty$, at time scales where the system has reached a steady state. Figure~\ref{fig:D_tilde_triggered} shows the fraction of cold particles in the largest cluster at steady state, $\Minfty / \Ncold$, as a function of $D$ for four packing fractions $\phi$. 
We find that for small $D$ almost all cold particles end up in the largest cluster. However, with increasing $D$ we find a pronounced drop in $\Minfty / \Ncold$, which sets in at a point that strongly depends on the packing fraction $\phi$. This suggests that there is a threshold value $\Dthresh (\phi)$, which marks a phase transition from an isotropic phase, where cold and hot particles are homogeneously mixed, to a demixed phase, in which they are segregated into a solid phase of cold and a gaseous phase of mainly hot particles.

\begin{figure}[t]
\centering
\includegraphics[width=0.9\columnwidth]{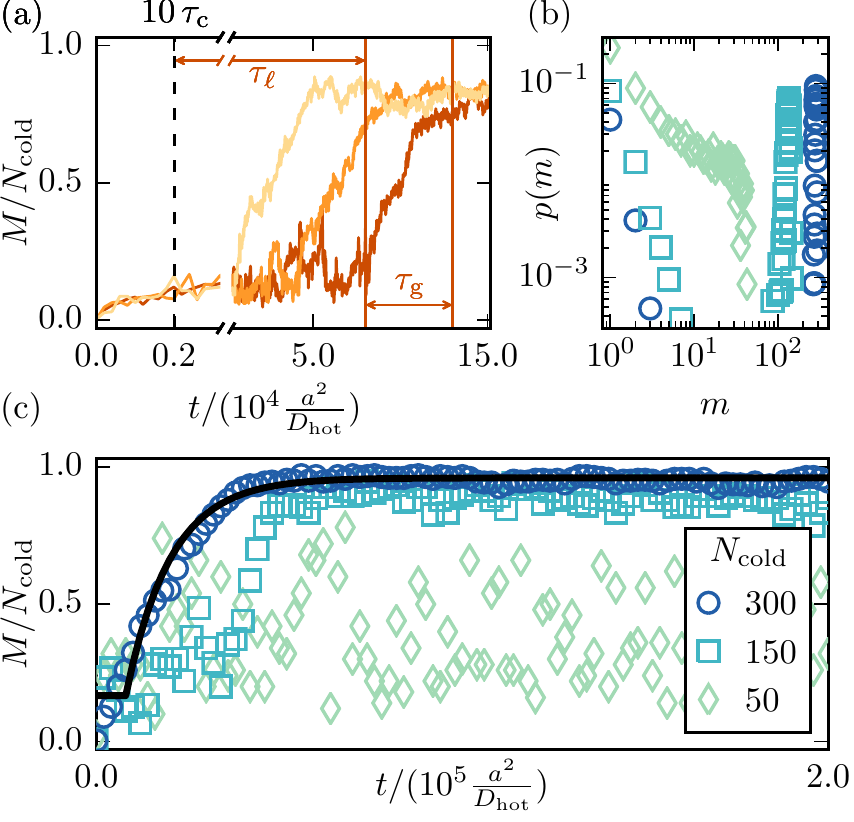}
\caption{%
\text{(a)} Time evolution of the fraction of cold particles in the largest cluster $M/\Ncold$ in a system of $\Nhot=\Ncold=300$ hot and cold particles with diffusion constant ratio $D = 1.5 \times 10^{-3 }$
 at total packing fraction $\phi=0.2$.  Each color (gray shade) depicts a single realization of initially randomly distributed particles. The lag time $\tlag$ and the time for cluster growth $\tgrowth$ is shown for the rightmost curve. The dashed line indicates ten collision times $\tcollision$.
\text{(b)} Cluster size distributions of cold particles for systems with $\phi=0.2$ and $D = 0$.
Different symbols and colors correspond to different particle numbers; see legend of (c). \text{(c)} Time traces of $M/\Ncold$ (same system as (b)). Black solid line:  Numerical solution of Eq.~\eqref{eq:rate_equation} with  parameters as in Supplemental Material~\cite{SM}, Fig.~S2, started after the corresponding lag time.
}\label{fig:nucleation}
\end{figure} 

 To further quantify this phase separation scenario we explored the dynamics of cluster formation and growth. 
Particle motions are determined by \emph{intrinsic} diffusion (with diffusion constant $\Dhot$ or $\Dcold$) and particle collisions. For $\Dhot \gg \Dcold$, movement of cold particles is primarily driven by collisions with hot particles, leading to an effective mesoscopic diffusion constant $D_\text{eff}$ of the cold particles. Our simulations show that for a single cold particle immersed in a bath of hot particles at packing fraction $\PackingFractionHot$, the effective diffusion constant can be approximated by $D_\text{eff} =\gamma  \PackingFractionHot \Dhot$ with $\gamma \approx \ch{0.28}$; see Supplemental Material~\cite{SM}, Fig.~S1. 
As illustrated in Fig.~\ref{fig:nucleation}(a), four qualitatively different regimes can be discerned for the phase separation process. Initially, on a time scale determined by the collision time $\tcollision =  L^2/(4\Ncold\Deff)$, there is fast assembly of many small clusters of cold particles. These small clusters coexist for an extended lag time $\tlag$ which typically is much longer than the collision time $\tcollision$. 
Subsequently, for the system sizes considered, we observe coarsening dynamics which finally leads to the formation of a single cluster whose growth saturates due to depletion of cold particles.  While the duration of the growth period $\tgrowth$ is similar for different realizations of the stochastic dynamics, the lag time $\tlag$ varies significantly.

The large variance in $\tlag$ suggests that the coarsening dynamics is preceded by a cluster nucleation phase, i.e.\  a cluster must first reach a critical size before it can stably grow. As this implies that no stable clusters can emerge in finite systems with particle numbers close to or below the number required to form a viable nucleus, we measured the weighted cluster size distribution $p(m) = m \cdot n_m / \Ncold$ ($n_m$: frequency of clusters of size $m$ of cold particles) for different particle numbers $N$ while keeping the parameters $\phi=0.2$, $D=0$ and $\Nhot/\Ncold=1$ fixed. We find that for small systems ($\Ncold = 50$), there is a broad distribution of cluster sizes originating from the continuous assembly and disassembly of clusters [Fig.~\ref{fig:nucleation}(b,c)], while for larger systems ($\Ncold = 150$), a single and stable cluster can form. We conclude that coarsening indeed requires the formation of a critical nucleus whose size is of the order of $\Ncold \approx 150$ for the parameters considered in [Fig.~\ref{fig:nucleation}(b,c)].

After nucleation, growth of the cluster is driven by 
a balance of attachment and detachment processes. To quantify these processes we develop a phenomenological mean-field theory for the dynamics of the cluster mass $M(t)$. Assume that the cluster is approximately spherical with radius $R$ and hexagonally packed with packing fraction $\eta = 0.9$ [see Fig.~\ref{fig:eye_catcher}(a), inset] such that $M a^2 \approx \eta \pi R^2$. In two dimensions, the flux of diffusing particles towards a sphere of radius $R$ scales as the diffusion constant times the area density of the particles and is independent of $R$~\cite{Krapivsky:2010ti}. Hence the rate of accretion of cold particles with effective diffusion constant $\Deff$ scales with $\omega_\text{att} (M) = \Deff (\Ncold-M) / L^2$, where $\Ncold-M$ is the number of cold particles outside the cluster.
Detachment may be mediated either through collisions of hot particles with the cluster surface or through intrinsic diffusion of cold particles.
Similarly to the case of cold particle attachment, the detachment rate due to collisions with hot particles scales with $\omega_\text{coll} = \Dhot \Nhot / L^2$, i.e.\ the flux of hot particles towards the cluster.
For the detachment rate due to intrinsic diffusion we take $\omega_\text{int} (M)= \Nsurface / (\PackingFractionHot \tint)$, where $\tint = a^2 / \Dcold$ is the time it takes a cold particle to leave the cluster due to its intrinsic motion and $\Nsurface=\pi R / a \approx \pi \sqrt{M/\eta}$ is the number of cold particles on the cluster surface.
Moreover, because hot particles block cold particles from diffusing away from the cluster we multiply the corresponding time-scale by $\PackingFractionHot$ to account for this caging effect.

The three processes combined lead to the time evolution of the cluster mass:
\begin{equation}\label{eq:rate_equation}
	\partial_t M=
		\alphap \, \omega_{\mathrm{att}}(M) - \alpham \, \omega_{\text{coll}} - \alphac \, \omega_{\text{int}}(M) \, ,
\end{equation}
where $\alphap$, $\alpham$ and $\alphac$ denote dimensionless coefficients, which were determined by fitting the solution of Eq.~\eqref{eq:rate_equation} to time-traces $M(t)$ obtained from ten independent realizations of our Brownian dynamics simulation for a single parameter set; see Supplemental Material~\cite{SM}, Fig.~S2 for details. We obtain $\alphap = \ch{82.5}$, $\alpham = 0.03$ and $\alphac = 7.5 \cdot 10^{-3}$. Using these values 
we find that the solutions to Eq.~\eqref{eq:rate_equation}
are in good agreement with the
simulation time-traces for different packing fractions or cold particle diffusion constants [Fig.~\ref{fig:nucleation}(c) and Fig.~S2 in the Supplemental Material~\cite{SM}].
Overall, this analysis shows that cluster growth is limited by collision-mediated diffusion of cold particles, and that cold surface particles are strongly attached to the cluster structure, with detachment events being rare. Moreover, the mean-field analysis also gives the correct saturation values for the cluster mass $\Minfty$  in the regime where a single and stable cluster develops [Fig.~\ref{fig:D_tilde_triggered}, shaded area]. As expected, outside of this regime Eq.~\eqref{eq:rate_equation} deviates from the simulations, since it does not take the coexistence of multiple clusters of fluctuating size and shape into account. A complete theory of cluster growth would need to incorporate cluster fragmentation and coalescence as well as Ostwald ripening~\cite{Kinetic_description_Oswald_el_driven_granules_2005}.

\begin{figure}[hbt]
\centering
\includegraphics[width=\linewidth]{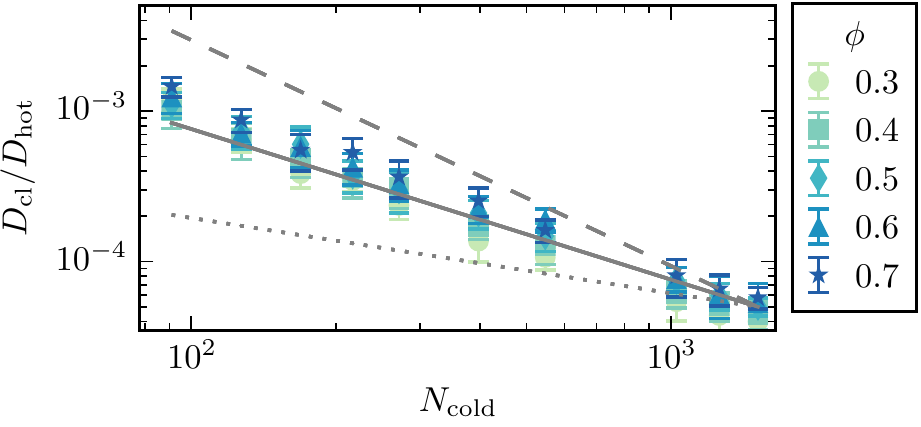}
\caption{% 
Diffusion constant for a saturated cluster of cold particles $\Dcl$ in units of $\Dhot$ as a function of the total number of cold particles $\Ncold$ for various packing fractions $\phi$ indicated in the graph; the intrinsic diffusion constant of cold particles was chosen as $\Dcold = 0$.
The scaling remains unchanged for $\Dcold>0$ (Supplemental Material~\cite{SM}, Fig.~S7).
 Error bars indicate one standard deviation of the distribution of $\Dcl$ values obtained from ten independent simulations. The three straight lines indicate power law scalings $\Dcl \sim \Ncold^{-\kappa}$:  $\kappa = 1.5$ (dashed), $\kappa = 1.0$ (solid), $\kappa = 0.5$ (dotted). The best fit was obtained for $\kappa = 1$.}\label{fig:cluster_diff}
\end{figure}

The observed saturation of cluster growth is a consequence of the finite numbers of cold particles in the systems considered in our computer simulations \ch{(see Supplemental Material~\cite{SM})}. However, for large systems, one expects that multiple clusters of cold particles should form throughout the system. Initially, they will grow due to accumulation of cold particles from the immediate vicinity as described above. 
At later stages, these clusters will however exhibit coarsening dynamics according to one or both of the following two principal mechanisms. Either there will be growth of larger clusters at the expense of smaller clusters (Ostwald ripening)~\cite{Lifshitz_Slyozov_61}, or clusters will meet by diffusion and then coalesce.
While our simulations do not allow us to draw any definite conclusion with respect to the impact of Ostwald ripening, we can estimate the coarsening dynamics due to cluster coalescence. To this end, we measured the diffusion constant of clusters consisting of cold particles as a function of their size. We find that the diffusion constant $\Dcl(\Ncold)$ of saturated clusters scales in inverse proportion to $\Ncold$ [Fig.~\ref{fig:cluster_diff}].
 In terms of cluster radius $R$ this implies that $\Dcl(R) \sim R^{-2}$, which suggests a `surface diffusion' mechanism for cluster diffusion~\cite{Khare:1995}: Single cold particles detach from the cluster surface very rapidly, then slowly diffuse away from the cluster surface and reattach to another surface site~\cite{Khare:1995, Morgenstern_2001}. This is consistent with our model lacking any explicit attractive interactions between cold particles and their slow effective diffusion. 
We found that a cluster moving via `surface diffusion' would grow through coalescence via $R(t)\sim t^{1/4}$; see Supplemental Material~\cite{SM}. 

\begin{figure}[t]
\centering	
\includegraphics[width=0.8\linewidth]{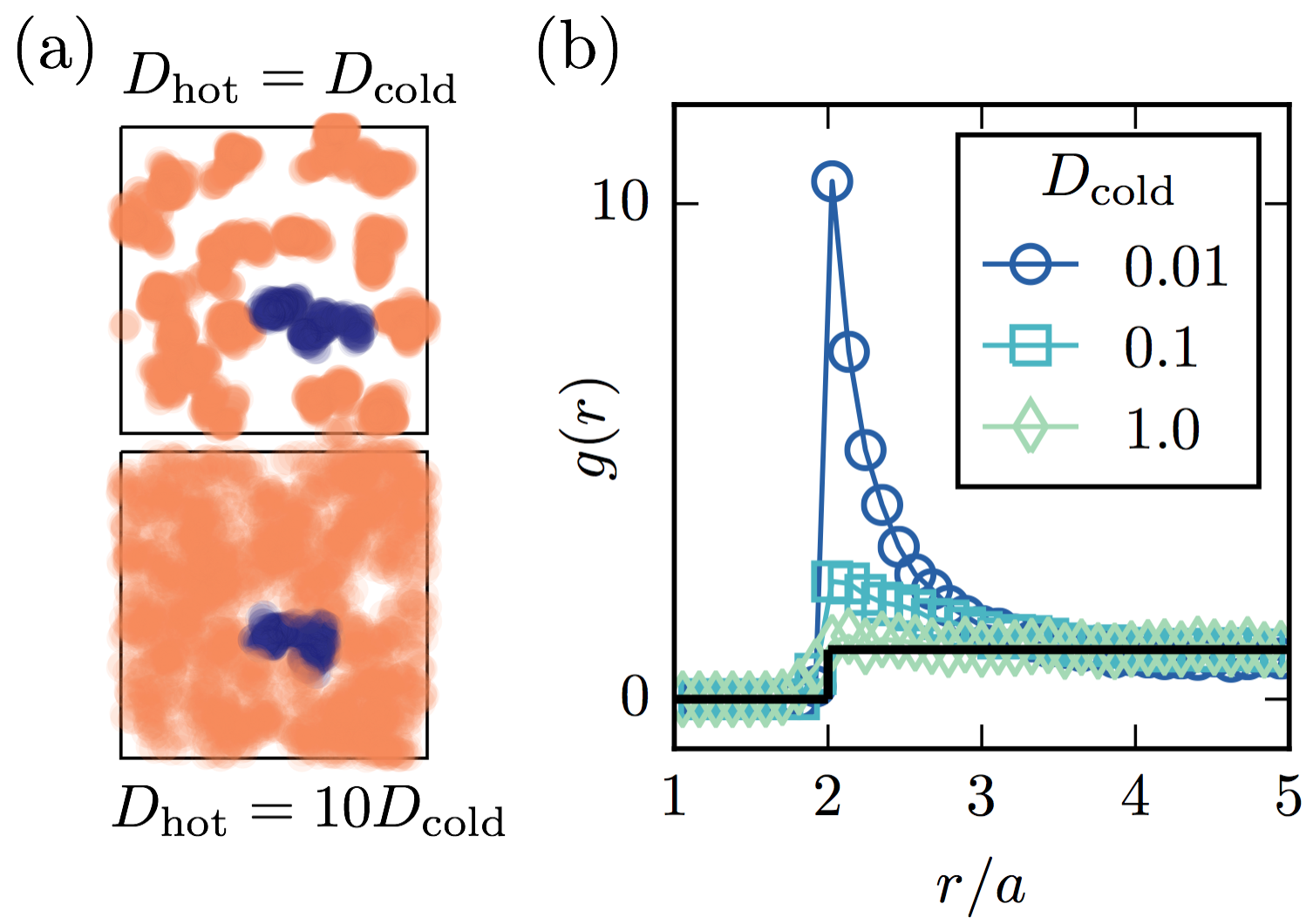}
\caption{%
\text{(a)} 
Illustration of caging. 
Two cold particles (blue, dark gray) are initially ($t=0$) in contact and homogeneously surrounded by hot particles (orange, light gray). 
The figure shows two simulations at time $t=2 a^2/\Dcold$ with the same diffusion constant of cold particles $\Dcold$. For each time increment of $0.01 a^2/\Dcold$ a transparent disk with radius $a$ is drawn, such that the density of disks depicts the positions of particles during the simulation.
Top: Simulation with equal diffusion constants $\Dhot=\Dcold$.
Bottom: Same simulation but with increased diffusion constant of hot particles $\Dhot=10\,\Dcold$.
\text{(b)}
Radial pair distribution function $g(r)$ for two cold particles surrounded by 11 hot particles at $\phi=0.13$ for three different diffusion constant ratios $D=\Dcold/\Dhot$. Solid black line:  $g(r)$ of hard disks in the dilute limit ($\phi \to 0$).
}
\label{fig:two_cold_particles}
\end{figure}

Finally, we asked why--despite  the lack of any explicit attractive interaction--there is any clustering of cold particles at all. 
To this end we considered two cold particles immersed in a `bath' of hot particles. 
As illustrated in Fig.~\ref{fig:two_cold_particles}(a), in any given time interval hot particles cover much longer distances than cold particles, and hence will frequently collide with the two cold particles. In particle configurations, where two cold particles are close together, these collisions will tend to keep them that way, a phenomenon reminiscent of high-density caging of hard disks~\cite{nijboer1952radialdistribution}.
To quantify the degree of caging created by hot particles we determined the radial distribution function $g(r)$, starting from a random configuration; see Fig.~\ref{fig:two_cold_particles}(b).
For $\Dcold \ll \Dhot$, $g(r)$ clearly shows that distances between cold particles are likely to be shorter than when  $\Dcold=\Dhot$, where $g(r)$ is essentially flat. We conclude that there is an effective attractive interaction between colder particles due to collisions with the surrounding hot particles which essentially act as a cage. This caging effect is fundamentally different from the well-known depletion interaction observed in colloidal suspensions containing particles with markedly different sizes \cite{PhysRevLett.83.3960, Marenduzzo:2006ch}. According to Asakura and Oosawa~\cite{Asakura:1958tn}, the effective interaction between the larger particles is mainly an entropic effect: When the large particles are close together, the regions inaccessible to the smaller particles overlap, leading to an increase in the entropy of the system.

Demixing in binary mixtures consisting of particles with different diffusion constants 
 is fundamentally different from phase separation in mixtures of 
 Brownian (passive) and self-propelled (active) particles, where the coexisting gas and solid-like phase contain both particle species~\cite{stenhammar2015activity}.
In these mixtures of active and passive particles, the gas-solid phase separation occurs at very high P\'eclet numbers $\text{Pe}= 3 \ell / (2 a)$ (where $\ell$ is the persistence length of the active particles) and even in the absence of passive particles~\cite{Fily:2012te,Redner:2013,Buttinoni:2013_sp}. 
In contrast, in our simulations the binary nature of the system is paramount since the \emph{differences} in activity between the species drive the phase separation. 
Moreover, the binary demixing discussed here occurs at $\text{Pe} = 0$. 
\ch{To compare our findings to the clustering at large P\'eclet numbers in active systems 
we also studied the case of non-zero P\'eclet numbers.
To this end we performed numerical studies of binary mixtures consisting of purely passive and active self-propelled particles without translational diffusion\ch{; note that translational noise is not a requirement for phase separation in simulations of Langevin equations \cite{fily2014freezing,levis2014clustering}.}
For the model and the results see Supplemental Material~\cite{SM}, Fig.~S4 and Fig.~S5.}
We find that demixing between cold and hot particles occurs even for $\text{Pe} > 0$.
However, increasing the P\'eclet number further, demixing vanishes, while for very large values, 
we observe coexisting gas and solid-like phases which contain both particle species, reminiscent of the findings in Ref.~\cite{stenhammar2015activity}.
This suggests that demixing of passive and active particles 
is driven by different mechanisms depending on the value of the P\'eclet number.

Our predictions for  binary mixtures consisting of particles with different diffusion constants can be readily tested for mixtures of granular disks or suspensions of
driven colloidal particles.
The key ingredient for each system is a local mechanism that enables the particles \ch{to move with different diffusion constants}.
In the case of vibrated colloids, a horizontal plate is attached to a shaker, which induces vertical and/or horizontal oscillations.
By capping the plate with a lid, one can restrict vertical motion sufficiently to effectively confine the colloids to two dimensions~\cite{Melby:2005gh}. 
Systems of two disk-like species of identical size but with different diffusivity can be implemented by using particles with differently profiled bases~\cite{Deseigne:2010collective, Deseigne:2011thesis, weber_disks_2013}, different materials~\cite{Rivas:2011, Rivas:2011ht} or, potentially, different heights.
\ch{It should be ruled out that the collision dynamics,  e.g.\ between particles of different weight, is the cause for clustering}.
Driven colloidal suspensions represent another possible realization of our model~\cite{bialke2015active,palacci2013living,Buttinoni:2013_sp}. For example, consider a system consisting of
synthetic photoactivated colloidal particles~\cite{palacci2013living} or \ch{non-isotropically} coated Janus particles~\cite{Buttinoni:2013_sp, howse2007self}, to which a second species is added that exhibits only Brownian motion. Given that the persistence length of the active species is less than its radius, this provides two species with different diffusion constants. 

\ch{
While our computational results show that differences in diffusion constant alone suffice to drive phase separation in binary mixtures, the phase behavior in actual  systems may be even more intriguing. 
It will be interesting to explore how differences in the degree of persistence in the particle trajectories--especially in the regime intermediate between the one studied here and in~\cite{stenhammar2015activity}--affect the systems dynamics and ensuing steady states.
Another promising route is to explore how the effective attraction between cold particles, mediated by caging through hot particles, is affected by additional interactions such as entropic forces mediated by differences in the size and shape of the particles.
}

\begin{acknowledgments}
This research was supported by the German Excellence Initiative via the program `NanoSystems Initiative Munich' (NIM), and the Deutsche Forschungsgemeinschaft (DFG) 
in the context of SFB 863 ``Forces in Biomolecular Systems" (Project B02).
\end{acknowledgments}

%\bibliographystyle{apsrev4-1}
%\bibliography{literatur}

%merlin.mbs apsrev4-1.bst 2010-07-25 4.21a (PWD, AO, DPC) hacked
%Control: key (0)
%Control: author (72) initials jnrlst
%Control: editor formatted (1) identically to author
%Control: production of article title (-1) disabled
%Control: page (0) single
%Control: year (1) truncated
%Control: production of eprint (0) enabled
%

\newpage\null\thispagestyle{empty}\newpage
%\newpage\null\thispagestyle{empty}\newpage
\begin{widetext}
%\newpage


\section*{Supplementary material}

# Supplemental Material:
# Binary Mixtures of Particles with Different Diffusivities Demix

Simon N. Weber,[1] Christoph A. Weber,[2] and Erwin Frey[1]

[1] *Arnold Sommerfeld Center for Theoretical Physics and Center for NanoScience,*
*Department of Physics, Ludwig-Maximilians-Universität München, 80333 Munich, Germany*
[2] *Max Planck Institute for the Physics of Complex Systems, Nöthnitzer Str. 38, 01187 Dresden, Germany*

## I. IMPLEMENTATION DETAILS OF THE NUMERICAL SIMULATIONS

In order to mimic hard particles, we use a large value for the product of spring constant and mobility $k\mu$. We choose $k\mu a^2/D_\text{hot} = 100$ with a time-discretization of $dt = 2 \times 10^{-3}$ implying typical overlaps of $\sqrt{2D_\text{hot}dt} \approx 0.06\, a$ which decay in a few time steps. We tested that the presented results are robust against reasonable changes in $k\mu$. As described in the main text and in Fig. S1, the effective diffusion constant of cold particles $D_\text{eff}$ can be estimated by $D_\text{eff} = \gamma \phi_\text{hot} D_\text{hot}$, with $\gamma \approx 0.28$. We simulate packing fractions $\phi = \phi_\text{hot} + \phi_\text{cold}$ between 0.2 and 0.7. So even at high packing fractions we get $D_\text{eff} < D_\text{hot}/10$. The relevant time scale for cluster growth is in the order of $(L/2)^2/D_\text{eff}$, whereas the numerical integration time step, $dt$, is limited by $a^2/D_\text{hot}$. The large difference in these two time scales makes simulations of larger systems computationally challenging.

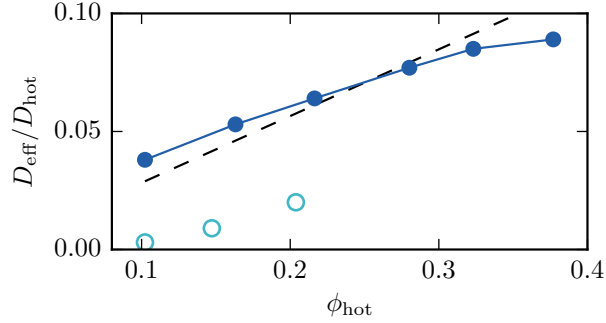

FIG. S 1. Effective diffusion constant $D_\text{eff}$ in units of $D_\text{hot}$ of a single cold particle in a bath of hot particles at packing fraction $\phi_\text{hot}$. The diffusion constant rises roughly linear with the packing fraction (dark blue circles). To extract this density dependence we fit a linear function $D_\text{eff} = \gamma \phi_\text{hot}$ to the data. Note that $D_\text{eff} = 0$ for $\phi_\text{hot} = 0$. The dashed black line shows a linear fit with slope $\gamma = 0.28$. Open circles depict the diffusion constant ratio $D = D_\text{cold}/D_\text{hot}$ at the transition points in Fig. 2, indicating that diffusion of cold particles is dominated by collisions with hot particles in the regime where phase separation occurs. For each packing fraction, the transition points are defined as the highest value of $D$ for which $M_\infty/N_\text{cold} \gtrsim 0.5$.

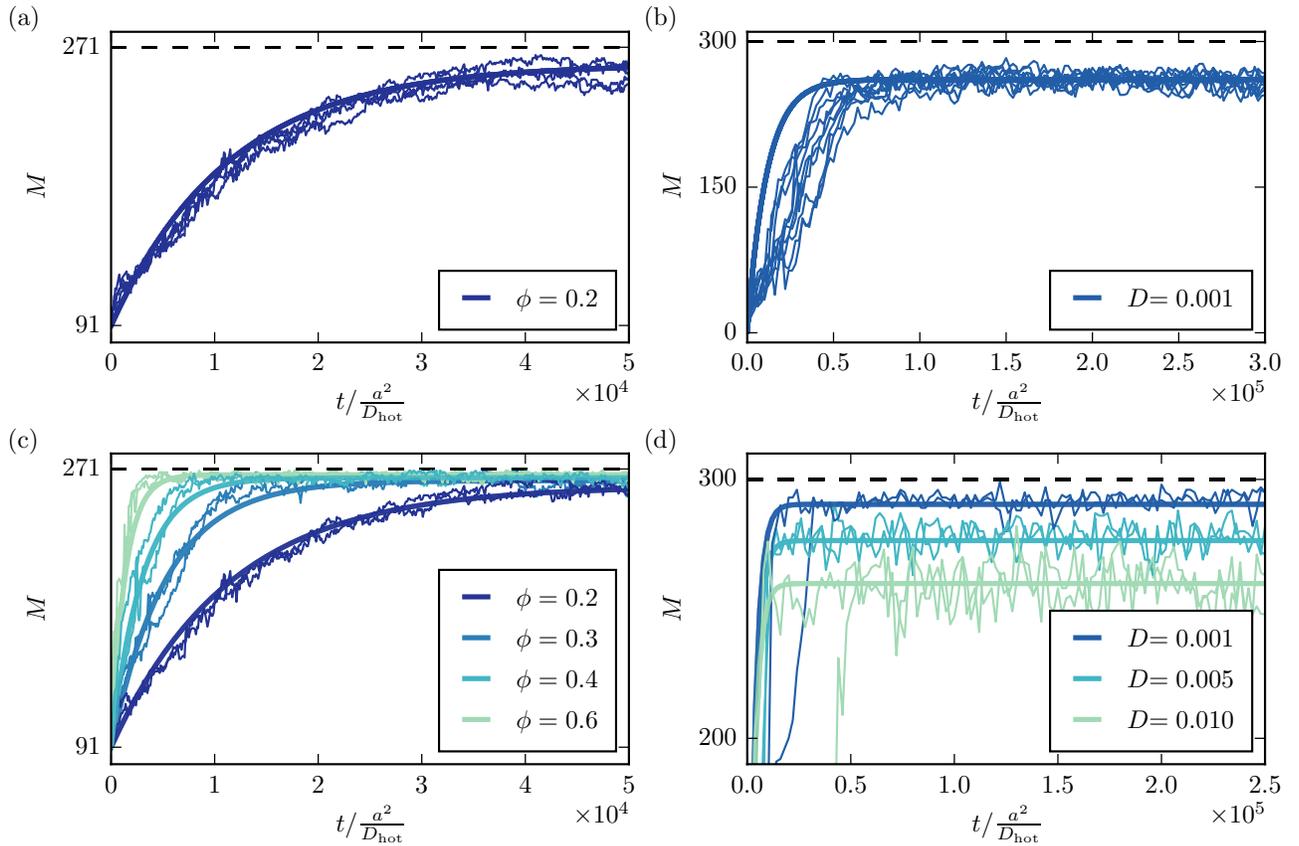

FIG. S 2. Finding the parameters $\alpha_{\text{att}}, \alpha_{\text{coll}}, \alpha_{\text{int}}$ through fitting. The time traces show the number $M$ of cold particles in the largest cluster of the system. For each parameter set $(D, \phi)$ a thick solid line shows the solution to Eq. (1) and thin solid lines depict two corresponding simulation results. **(a)** The parameters $\alpha_{\text{att}}, \alpha_{\text{coll}}$ for the attachment and detachment of cold particles through collisions with hot particles are determined by fitting to cluster growth curves with $D_{\text{cold}} = 0$. To avoid multi-site growth, 91 out of $N_{\text{cold}} = 271$ cold particles are initially hexagonally close-packed. All remaining particles are distributed randomly. We fitted $\alpha_{\text{att}}, \alpha_{\text{coll}}$ to time courses at a single packing fraction $\phi = 0.2$ and obtained $\alpha_{\text{att}} = 82.5, \alpha_{\text{coll}} = 0.03$. **(b)** To obtain the parameter for particle detachment due to intrinsic diffusion, $\alpha_{\text{int}}$, we use the parameters $\alpha_{\text{att}}$ and $\alpha_{\text{coll}}$ as before and simulate a system with $D = 0.001$, again at packing fraction $\phi = 0.2$. Choosing $\alpha_{\text{int}} = 7.5 \cdot 10^{-3}$ gives the correct cluster size at late times. **(c)** Similar simulations as in (a) but for different packing fractions $\phi$. We use the parameters $\alpha_{\text{att}}, \alpha_{\text{coll}}, \alpha_{\text{int}}$ as in (a) and (b). The solution of Eq. (1) is still in good agreement with the simulation results for different packing fractions. **(d)** Simulation as in (b) but with packing fraction $\phi = 0.4$ and varying diffusion constant ratios $D$. Again we keep the fitting parameters fixed to the values obtained in (a) and (b). The same set of parameters describes the cluster growth at different densities and different $D$. Please note that we keep the values for all three fitting parameters $\alpha_{\text{att}}, \alpha_{\text{coll}}, \alpha_{\text{int}}$ fixed in every plot in the Supplemental Material and the main text.

## II. RIPENING TIME EVOLUTION THROUGH COALESCENCE

In general two ripening mechanisms exist: Coalescence of clusters and the growth of large clusters at the expense of smaller clusters, referred to as Ostwald-ripening [1]. Here, we ignore Ostwald-ripening in our argument and restrict ourselves to giving a minimal proof for ripening by only considering coalescence of clusters. Say $\Delta$ denotes the inter-particle cluster distance and $\delta t$ the time to collide with another cluster, $\delta t \sim \Delta^2/D_{\rm cl}$. $D_{\rm cl}(R)$ denotes the diffusion constant of the cluster, which depends on the cluster radius $R$, $D_{\rm cl}(R) \approx C_0/R^2$ (Fig. 4); $C_0$ only marginally depends on packing fraction thus we treat it as constant. In order to obtain a scaling for the increase of the cluster radius with time, we consider an arrangement of boxes of side length $L$; see Fig. S3 for an illustration. Each box contains a single cluster of radius $R$ that covers an area fraction $\phi_A = \pi R^2/L^2$; similar to the one depicted in Fig. 1(b). This implies a mean inter-cluster distance of $\Delta \approx L = R\sqrt{\pi/\phi_A}$. A collision between clusters is assumed to conserve the overall cluster area fraction $\phi_A$ neglecting fragmentation events during the collision process. Therefore, the incremental radius change during a collision is given by $\delta \ln R(t) = \ln(2)/2$ since $R + R \to \sqrt{2}R$. This leads to $\frac{\delta}{\delta t} \ln R(t) = R^{-4} \phi_A C_0 \ln(2)/(2\pi)$. Integration yields the scaling $R(t) \sim t^{1/4}$, which suggests that the system ripens to a single cluster due to coalescence.

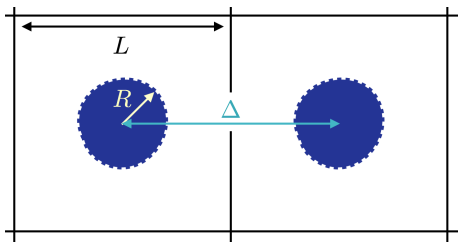

FIG. S 3. Illustration of the quantities used to derive the scaling argument for ripening through coalescence. The figure shows two compartments of an infinite system. In each compartment a cluster of radius $R$ has formed. The cluster distance $\Delta$ is approximated by the compartment size $\Delta \approx L$.

## III. MODEL FOR ACTIVE AND PASSIVE PARTICLES

Under which circumstances do binary mixtures of self-propelled particles and purely passive particles demix due to the same mechanisms that we observe in mixtures of particles with different diffusivities? To address this question we modeled mixtures of active and passive particles. Active particles perform self-propelled motion whose direction is subject to diffusion. Passive particles only move through collisions with hot particles. We do not add translational diffusion to the model. Physically, this means that we assume translational and rotational diffusion to be independent and translational diffusion to be negligibly small, so that the limit of small persistence lengths effectively results in the model described in the main text with $D = 0$. This assumption is only valid for systems where translational and rotational diffusion are not coupled. Note that in this respect the following model differs from the model described in [2]. However, as shown in [3], and further explained in [4], translational noise is not crucial for phase separation in simulations of Langevin equations. The dynamics for active and passive particles is governed by the following set of equations:

$$\partial_t \mathbf{r}_i = \mu \sum_{j=1}^{N} \mathbf{F}_{ij}(t) + v_i \begin{pmatrix} \cos(\theta_i(t)) \\ \sin(\theta_i(t)) \end{pmatrix} \,, \quad \partial_t \theta_i = \eta_i^{\rm rot}(t) \,, \quad i \in \{1, \ldots, N\} \,. \qquad (1)$$

There are $N = N_{\rm a} + N_{\rm p}$ particles in the system, where $N_{\rm a}$ and $N_{\rm p}$ are the number of active and passive particles. $\mathbf{F}_{ij}$ denotes the same repulsive force as described in the main text. The particles differ only in their propulsion speed: $v_i = v$ for active particles and $v_i = 0$ for passive particles, i.e. passive particles are not propelled and move only by collisions with active particles. To keep particle overlaps small we choose $\mu k a/v = 100$. The rotational noise on the polar vector $\theta_i$ is uniform and uncorrelated: $\eta_i^{\rm rot}(t) = \frac{2v}{\ell}\delta_{ij}\delta(t-t')$, where $\ell$ denotes the persistence length of the active particles. The smaller the value $\ell$ the more diffusive the motion of active particles. We find that for $\ell/a \leq 1$ passive particles form large clusters [Fig. S4] similar to the clusters of cold particles in mixtures of hot and cold particles in the main text. At these small persistence lengths we do not observe a clustering of active particles, which would be the case for large persistence lengths (see Fig. S5 and [2, 5]). This indicates that the effect of persistent motion disappears before the effect of diffusive motion becomes dominant.

4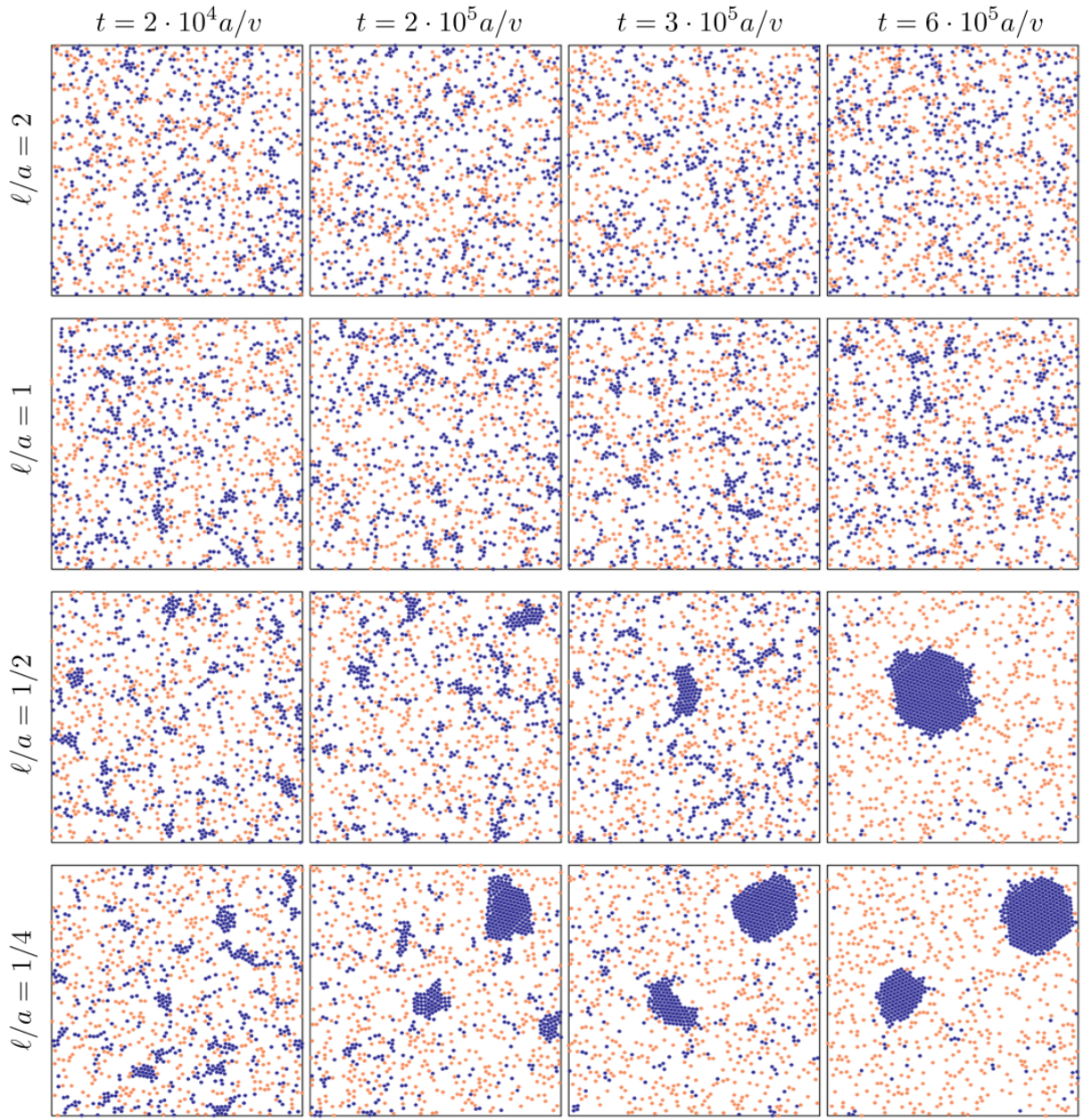

FIG. S 4. Systems of passive and self-propelled particles at small persistence lengths behave similarly to our model. Depicted are snapshots for simulations with $N_\text{a} = N_\text{p} = 547$ active (orange) and passive particles (blue), respectively. The total packing fraction is $\phi = 0.2$. Different rows correspond to different persistence lengths $\ell$ and different columns to different moments in time. For each system the particles were initially randomly distributed. For $\ell/a \geq 1$ only small clusters emerge. For $\ell/a < 1$, i.e. effectively diffusive motion, the passive particles assemble in large clusters as in simulations with particles of different diffusivity, where the cold particles cluster.

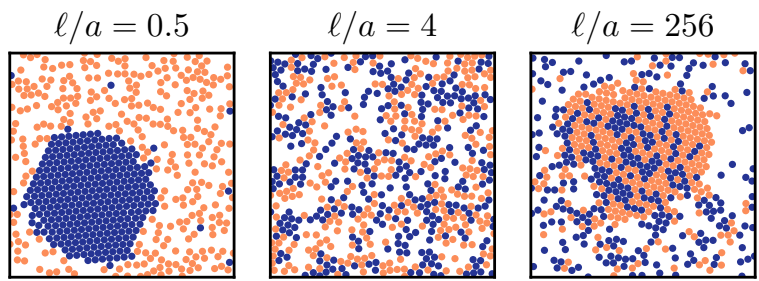

FIG. S 5. The persistence length of the active particles determines the type of clustering. Shown are snapshots at late times of mixtures of active (orange) and passive (blue) particles. Initially the particles are distributed at random. For short persistence lengths $\ell < a$, i.e. effectively diffusive motion, passive particles form a solid cluster. For intermediate persistence lengths $\ell \approx a$ no large clusters emerge. For large persistence lengths $\ell \gg a$ the active particles form a jammed cluster that also contains passive particles. This scenario corresponds to large Péclet numbers in [2].

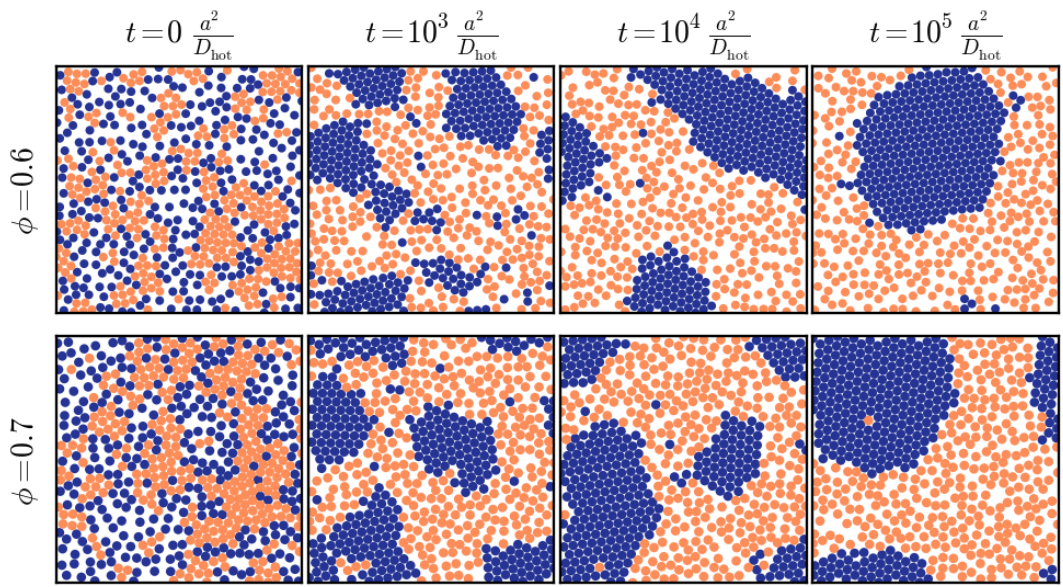

FIG. S 6. Clustering at high packing fractions. Snapshots for packing fraction $\phi = 0.6$ (top) and $\phi = 0.7$ (bottom) at different moments in time. Eventually a single large cluster of cold particles emerges from an initially homogeneous distribution.

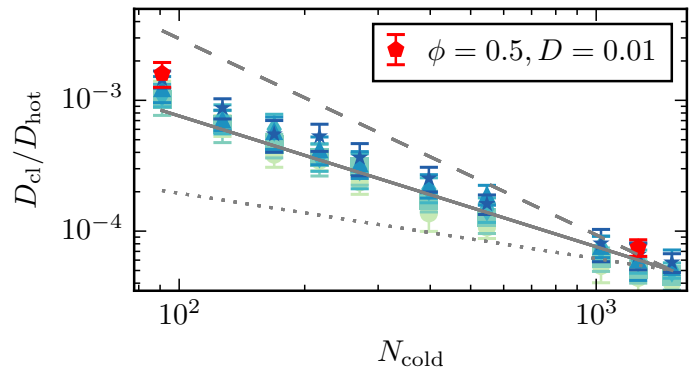

FIG. S 7. Diffusion constants for saturated clusters $D_{\text{cl}}$ as in Fig. 4 in the main text. Red pentagons show the diffusion constant for saturated clusters in systems with packing fraction $\phi = 0.5$ and diffusion constant ratio $D = 0.01$ of cold and hot particles. Even at this high value for the intrinsic diffusion constant of cold particles the scaling remains $D_{\text{cl}} \sim N_{\text{cold}}^{-1}$ which again corresponds to surface diffusion (see main text). Note that for $D > 0$ the values for $D_{\text{cl}}$ are slightly higher than for $D = 0$. This makes sense, because intrinsic diffusion of cold particles speeds up the process of surface diffusion.

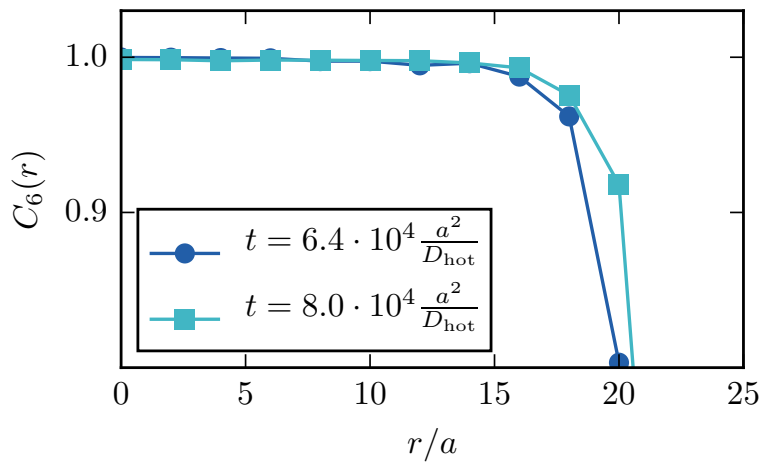

FIG. S 8. The equal-time spatial correlation function $C_6(r) = \frac{1}{\sum_i |\Psi_{6,i}|^2} \sum_{|\mathbf{x}_i - \mathbf{x}_j| = r} \Psi_{6,i} \Psi_{6,j}^*$ for the simulations shown in Fig. 1 in the main text at two different moments in time. For the reference position $\mathbf{x}_i$ we take the position of the cluster particle which is closest to the center of mass of the cluster. $C_6$ starts to decrease only at the edge of the cluster. This indicates that the cluster is hexagonally arranged in its entire bulk.

## IV. VIDEO

The video corresponds to the simulation shown in Fig. 1 in the main text. Initially both particle species are randomly distributed. Many shapeless clusters of cold particles emerge quickly. At longer times, a large cluster forms that contains most of the cold particles. This cluster remains stable. Time is measured in units of $a^2/D_{\text{hot}}$. Parameters: $N_{\text{hot}} = N_{\text{cold}} = 500, D = 10^{-3}, \phi = 0.2$.

 
\end{widetext}

\end{document}